\documentclass[aps,prb,twocolumn,showpacs,superscriptaddress]{revtex4}
\usepackage{graphicx}
\usepackage{hyperref}
\usepackage{natbib}
\begin{document}
\title{Long-lived memory for electronic spin in a quantum dot: Numerical
analysis}
\author{V. V. Dobrovitski}
\affiliation{Ames Laboratory, Iowa State University, Ames, IA 50011, USA}
\author{J. M. Taylor}
\affiliation{Department of Physics, Harvard University, Cambridge, Massachusetts 02138, USA}
\author{M. D. Lukin}
\affiliation{Department of Physics, Harvard University, Cambridge, Massachusetts 02138, USA}
\begin{abstract}
  
  Techniques for coherent control of electron spin-nuclear spin
  interactions in quantum dots can be directly applied in spintronics and in
  quantum information processing.  In this work we study numerically
  the interaction of electron and nuclear spins in the context of
  storing the spin-state of an electron in a collective state of
  nuclear spins.  We take into account the errors inherent in a
  realistic system: the incomplete polarization of the bath of nuclear
  spins and the different hyperfine interactions between the
  electron and individual nuclei in the quantum dot.  Although these
  imperfections deteriorate the fidelity of the quantum information
  retrieval, we find reasonable fidelities are 
  achievable for modest bath polarizations.
\end{abstract}
\pacs{03.67.-a,75.40.Mg,73.21.La,76.70.-r}

\maketitle

\section{Introduction}
Currently, the development of techniques for coherent manipulation and
control of electron spins in semiconductor nanostructures is an area
of vibrant research activity \cite{loss98,imamoglu99,%
  bracker04,elzerman04,johnson05,petta05}. Such techniques are needed
for many coherent spintronic devices \cite{awschalom01}, such as a
Datta-Das transistor \cite{datta90}, and for quantum computing using
the spin of an electron in a quantum dot (QD) as a qubit
\cite{loss98,imamoglu99}.  However, the decoherence time of an
electron spin constitutes a fundamental obstacle on the way to
realization of a QD-based quantum computation
\cite{prokstamp,dkdh}. In order to enhance the coherence time, a
quantum memory protocol has been recently suggested \cite{taylor03}
which allows mapping of the quantum state of the electron spin on the
collective spin of the nuclei present in the QD, thereby taking
advantage of the intrinsically longer coherence times of nuclear
spins. Under ideal circumstances, the proposed scheme allows to encode
and, at later time, to retrieve, the electron spin state with
near 100\% reliability. Furthermore, it can be extended to
implementing arbitrary 1- and 2-qubit operations \cite{taylor04}.
Imperfections, which should be expected in realistic settings, lead to
errors, thus deteriorating the fidelity of the quantum memory protocol
\cite{taylor03,deng04}.  Some errors can be reduced
\cite{taylor03b,giedke05}, but some are inherent to the protocol, such as the
incomplete polarization of the nuclear spins and the spread in the
hyperfine interactions between the electron and the nuclei in the QD.
In this work, we use numerical simulations
\cite{dobrovitski03b,deraedt04} to analyze the effect of these errors
on the performance of the quantum memory protocol. We show that the
scheme remains feasible for the spin bath polarizations exceeding
80\%, with the minimum fidelity of $\sim 80$\%.

First, let us recall the scheme of Ref.~\onlinecite{taylor03}.  The
electron with a spinor wave function $|\phi_0\rangle =
\alpha|\downarrow\rangle + \beta|\uparrow\rangle$ is injected into an
empty quantum dot, and starts interacting with the bath of $N$ nuclear
spins in the dot. The most important part of this interaction is the
isotropic contact hyperfine coupling.  The total Hamiltonian of the
system, taking into account the Zeeman energy of the electron spin in
the external field $H_0$, becomes
\begin{equation}
{\cal H} = g_e^*\mu_B H_0 S^z + \sum_{k=1}^{N} A_k {\mathbf I}_k{\mathbf
  S}
\label{hamt}
\end{equation}
where $S$=1/2 is the electron spin, $I_k=3/2$ are the spins of the
nuclei (the fact that the nuclear spins of Ga and As are not 1/2 is
not important here, up to renormalization of the appropriate
parameters, and we will consider them as spin-1/2 in the following
discussion), and $A_k=(8\pi/3)g_e\mu_Bg_n\mu_n |\Psi({\mathbf
  x}_k)|^2$ 
is the contact coupling determined by the electron density
$|\Psi({\mathbf x}_k)|^2$ at the site ${\mathbf x}_k$ of $k$-th bath
spin, and by the Land\'e factors $g_e$, $g_n$ and the magnetons
$\mu_B$, $\mu_n$ of the electron and the nuclei, respectively.  At
maximum polarization, the nuclei can produce an effective field
$\sum_k A_k I_k / g^*_e \mu_B = 5.2$ Tesla \cite{paget}.

Other terms in the Hamiltonian (\ref{hamt}) can be omitted when
discussing the storage and retrieval processes
due to their smallness at the relevant timescales. These are,
e.g., the Zeeman energy of the nuclear spins, the 
dipolar coupling between the bath spins, or between the electron
spin and the bath spins. Also, we assume that the injection
is performed quickly, so that the approximation of the sudden
change of the Hamiltonian is applicable, and neither the state of the
bath nor the state of the electron spin are affected by the
injection process.  Alternatively, fast control of the Zeeman
splitting of the electron allows for the electron spin-nuclear spin
system to be brought into resonance in a sudden approximation, and is
similar in nature to the controlled injection/ejection of electrons from the
quantum dot.

If the electronic wave function were spread uniformly over the dot
volume, making all $A_k$ equal to $A$, and if the nuclear spins were
completely polarized (i.e., if the bath were in the state
$|\chi_0\rangle = |\downarrow,\downarrow,\dots\downarrow\rangle$)
then, for the external field adjusted to be $H_0=A(N-1)/(2 g_e\mu_B)$,
the motion of the compound system (the electron and the bath spins)
would become a sinusoidal oscillation between the states
$|\phi_0\rangle\otimes|\chi_0\rangle$ and
$|\downarrow\rangle\otimes[\alpha|\chi_0\rangle +
i\beta|\chi_1\rangle]$ where $|\chi_1\rangle = (1/\sqrt{N})\sum_k
S_k^+ |\chi_0\rangle$ ($N$ is the number of the nuclei in the dot).
Thus, after half-period of such oscillations, the initial state of the
electron spin becomes encoded in the state of the bath, and the
electron can be ejected from the quantum dot. In order to retrieve
this information after some time, another electron in the state
$|\downarrow\rangle$ can be injected, and after a half-period of the
oscillations, the state of the electron spin is again $|\phi_0\rangle
= \alpha|\downarrow\rangle + \beta|\uparrow\rangle$.  Of course, for
this two-step scheme to work, the delay between the information
encoding and retrieval should be small in comparison with the
characteristic times of the omitted terms in the Hamiltonian (in
practice, these terms should be reduced using, e.g., NMR techniques
\cite{whh,vandersypen04}).

However, real baths are incompletely polarized and the approximation
of complete polarization is {\it never\/} applicable to a real bath
with $N\sim 10^4-10^6$. To explain this, let us consider a highly
polarized state of the bath described by a density matrix
$\rho=(1/Z)\exp{(-\gamma I^z)}$, where $I^z=\sum_k I_k^z$, and $Z$ is
chosen such that ${\rm Tr} \rho = 1$.  For such a bath, the polarization is
$P=\tanh{\gamma/2}$, and for large $\gamma$ the value of $\Delta P=1-P$
is very small. However, the statistical weight $w(M)$ of the states
with a given value $M$ of $I^z$ is $w(M) = C_k^N \theta^k
(1-\theta)^{N-k}$, where $\theta=e^\gamma/(1+e^\gamma)$, the
integer $k=N/2-M$, and $C_k^N$ is a binomial coefficient. 
The distribution $w(M)$ is approximately Gaussian
centered at ${\bar M}= (-N/2)(1-\Delta P)$ with the rms
$\sigma^2=(N/4)\Delta P(2-\Delta P)$. Therefore, unless $\Delta P$ is
unrealistically small ($\Delta P\ll 1/\sqrt{N}\sim 10^{-2}$), a very
large number of the bath states are involved in the time evolution, as
opposed to only two states $|\chi_0\rangle$ and $|\chi_1\rangle$
involved in the case of $A_k=A$ and $\Delta P=0$. All these states have
different oscillations frequencies, and form complex time-dependent
superpositions, so that the fidelity of the information encoding and
retrieval is reduced.  Therefore, the errors associated with
$\Delta P>0$ (incomplete polarization) and $A_k\neq A$ (spread in the
hyperfine couplings) are intrinsic to the protocol. Curiously, the symmetry
properties of the spin interaction allows for some correction, by
state engineering and spin-echo techniques, for these errors
\cite{taylor03b}.  

The paper is organized as follows.  In section II, we consider
qualitatively the analytical behavior of the storage protocol when
inhomogeneity is negligible ($A_k = A$) but polarization is less than
perfect, and indicate how many of the effects of
different oscillation frequencies and Hilbert space dimension lead to
only moderate decreases in the fidelity of storage and retrieval.  
We then evaluate the same situation numerical in section
III, and develop the useful measures of performance for the protocol.
In section IV, the numerical results are extended to the case of
inhomogeneous interaction ($A_k \neq A_{k'}$), and then in section V
the minimal fidelity of the protocol (minimized over all possible
electron spin states) is calculated.

\section{Qualitative analytics for $A_k=A$: Why the protocol works
for $1/\sqrt{N}\ll\Delta P\ll 1$}
\label{QualitAnalyt}

As indicated above~\cite{taylor03}, for $A_k=A$ and
$\Delta P>0$ the fidelity of the quantum memory protocol remains high
if $\Delta P\ll 1$, even for $\Delta P\gg 1/\sqrt{N}$.
It is instructive to understand this robustness in detail.

To grasp the idea, let us consider only the first, encoding, step.
We note that for $A_k=A$, the total spin of the nuclear bath $I$
is an integral of motion. If the initial state of the
bath is $|I_0,M_0\rangle$, where $I_0$ is the value of $I$ and
$M_0$ is the value of $I^z$, then the time-dependent
wave function of the compound system is
\begin{eqnarray}
\label{oscills}
|\psi(t)\rangle &=& \beta|\psi_1(t)\rangle + \alpha|\psi_2(t)\rangle\\
\nonumber
|\psi_1(t)\rangle &=& (\cos{\omega_1t} - i\xi_1\sin{\omega_1t})
  |\uparrow\rangle\otimes|I_0,M_0\rangle \\
\nonumber
  &-&i\eta_1\sin{\omega_1t}|\downarrow\rangle\otimes|I_0,M_0+1\rangle\\
\nonumber
|\psi_2(t)\rangle &=& (\cos{\omega_2t} + i\xi_2\sin{\omega_2t})
  |\downarrow\rangle\otimes|I_0,M_0\rangle \\
\nonumber
  &-&i\eta_2\sin{\omega_2t}|\uparrow\rangle\otimes|I_0,M_0-1\rangle
\end{eqnarray}
where $\omega_1 = \sqrt{h_1^2+A^2C_1^2/4}$, $\xi_1=h_1/\omega_1$,
and $\eta_1 = AC_1/(2\omega_1)$ ($\omega_2$, $\xi_2$ and $\eta_2$ are
defined similarly). The relevant matrix elements are
\begin{eqnarray*}
h_1 &=&g_e\mu_B H_0/2 + A(M_0+1/2)/2\ , \\h_2 & = & g_e\mu_B H_0/2 +
A(M_0-1/2)/2\ , \\
C_1 & = & \sqrt{(I_0-M_0)(I_0+M_0+1)}\ , \\C_2 &=
&\sqrt{(I_0+M_0)(I_0-M_0+1)}\ .
\end{eqnarray*}

The oscillation frequencies $\omega_{1,2}$ are strongly dependent on
both $I_0$ and $M_0$. For the bath's initial density matrix 
$\rho=(1/Z)\exp{(-\gamma I^z)}$, the statistical weight for the
state $|I_0,M_0\rangle$ is
\begin{equation}
w(I_0,M_0)=w(M_0) (C_m^N-C_{m-1}^N)/C_k^N
\label{wIM}
\end{equation}
where $k=N/2-M_0$ and $m=N/2-I_0$ are integers, 
and $C^N_k$ is a binomial coefficient. For $\Delta P\ll 1$, both
$N-k$ and $m$ are small (note that $M_0<0$, so that $k$ is close to $N$),
and for a given $M_0$, 
it is easy to see that $w(I_0,M_0) \sim \Delta P^{(N-k-m)}$.
The main contribution to the evolution comes only from the states with
$I_0=-M_0$, and the states with larger values of $I_0 ( > -M_0 )$ have too
small a weight to impact the oscillations, and, correspondingly, to
reduce the fidelity. Thus, the distribution of the frequencies
$\omega_{1,2}$ in Eq.~\ref{oscills} is determined only by $w(M_0)$,
and a spread of the oscillation frequencies $\sigma\sim \sqrt{N\Delta P}$ 
is much smaller than the mean frequency ${\bar\omega}\sim \sqrt{N}$.

Using Eqs.~\ref{oscills},\ref{wIM} it is easy to obtain an analytical
approximation for $N\gg 1$ and $1/\sqrt{N}\ll\Delta P\ll 1$ in
the leading order in $\Delta P$ (but uniformly in time), since
$\xi_1, \eta_2 \sim 0$, $\xi_2, \eta_1 \sim 1$. For the spin initially
polarized along the $x$-axis ($\alpha=\beta=1/\sqrt{2}$),
\begin{equation}
s_x(t)=\cos{\left(\frac{\tau}{2}\sqrt{N(1-\Delta P)}\right)}
  \exp{(-\frac{\sigma^2\tau^2}{8})}+O(\Delta P)
\end{equation}
while for the initial "spin-up" state,
\begin{equation}
s_z(t)=\cos{\left(\frac{\tau}{2}\sqrt{N(1-\Delta P)}\right)} + O(\Delta P)
\end{equation}
where $\tau=At$, and only $s_x(t)$ is influenced in the leading order
in $\Delta P$.  The difference between the results for transverse (x)
and parallel (z) states indicates that the memory protocol is not
isotropic: its performance depends in part upon the spin state being
stored. As such, in analyzing errors, we will consider several
different performance measures, described in detail below.  We first
consider the reliability of numerical simulations.

\section{Quantitative analytics for $A_k=A$ and reliability of
computations for small $N$}

We focus on the case of $A_k=A$ in order to assess reliability of our
simulations with $N=20$ (presented below, performed for the case of
$A_k\neq A$), and to quantify the fidelity in detail, without omission
of higher-order terms in $\Delta P$.  The case of $A_k=A$ is a good
check since it allows analytical solution, so that the results for
$N=20$ can be directly compared with the results for realistic value
$N\sim 10^4$.

\begin{figure}
\includegraphics[width=8cm]{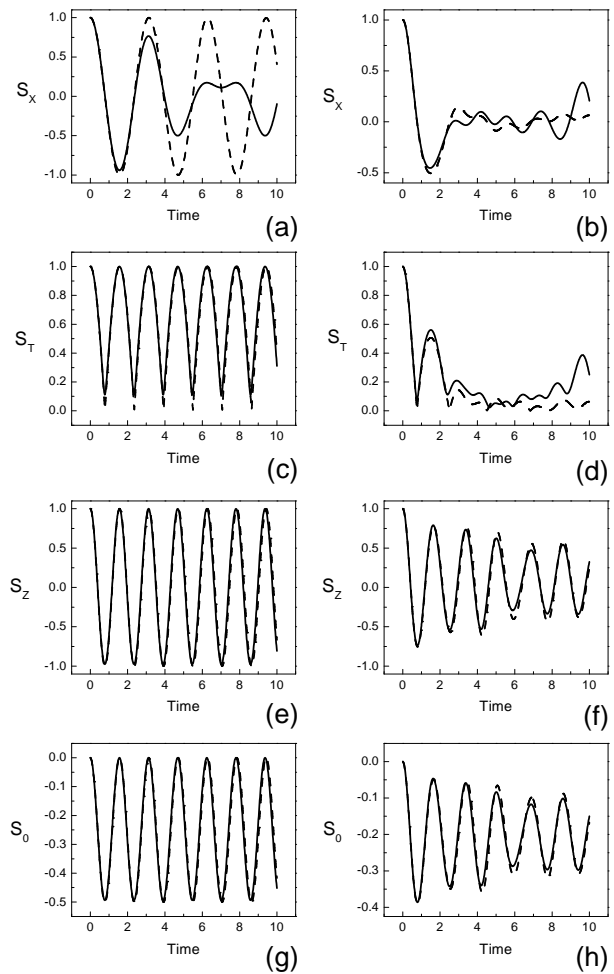}
\caption{First step of the protocol, encoding. Oscillations
are shown for $s_x$ (a,b), $s_T$ (c,d), $s_z$ (e,f) and $s_0$
(g,h), obtained for $\Delta P=0$ (a,c,e,g) and for 
$\Delta P=0.2$ (b,d,f,h), with $N=20$ (solid curves) and
$N=10^4$ (dashed curves).}
\label{fig1}
\end{figure}

Let us focus first on the encoding step to determine what will be
useful performance measures for the protocol.
Since $h_1\neq h_2$ in Eqs.~\ref{oscills},
the components $S^x$, $S^y$ of the electron spin rotate with time in
the $x$--$y$ plane with the angular velocity of order of $A/4$.  This
rotation is uniform, and does not depend on the particular values
$s_x$, $s_y$ of $S^x$ and $S^y$, because the Hamiltonian (\ref{hamt})
is invariant with respect to the rotation in the $x$--$y$ plane 
($\cal
H$ commutes with $\exp{[-i\epsilon(S^z+I^z)]}$ for any $\epsilon$).
As a result, even in perfect situation of $\Delta P=0$, $A_k=A$, for the
initial value of $s^{(i)}_y=0$, the final retrieved value
$s^{(f)}_y\neq 0$, and $s^{(f)}_x\le s^{(i)}_x$. However, this error is
noticeable only for relatively small $N$ ($N\le 100$), since the total
time of the interaction of the electron spin with the bath is only one
oscillation period, i.e., roughly $1/(A\sqrt{N})$, while the angular
velocity is of order of $A/4$. Furthermore, this error is easily
recoverable by application of a correcting pulse of the external field
$H_z$. Therefore, a physically meaningful measure of errors
associated with decoherence of the $S^x$ operator is the total
transverse component of the electron spin,
\begin{equation}
s_T=\sqrt{s_x^2+s_y^2}\ ,
\end{equation}
which is proportional to $s^{(i)}_x$
if $s^{(i)}_y=0$.  
This is illustrated in Figs.~\ref{fig1}(a--d), where we plotted
the oscillations of $s_x$ and $s_T$ for $\Delta P=0$ (left column,
(a) and (c)), and for $\Delta P=0.2$ (right column,(b) and (d)),
for the initial condition $s^{(i)}_y=s^{(i)}_z=0$, $s^{(i)}_x=1$.
The solid lines correspond to $N=20$, and the dashed lines
correspond to $N=10^4$. It is important that during the first
oscillation period, the results for $s_T$ obtained with 
$N=20$ are close to those obtained with $N=10^4$, so that
$N=20$ is sufficient to reliably estimate (with the precision
of $\sim 0.05$) the fidelity of
the quantum memory protocol at much larger values of $N$.  $s_T$ is one of the three performance measures we
use for the remainder of the paper.

In addition to the transverse errors, the oscillations of the value
$s_z$ of $S^z$ are {\it not simply\/} proportional to the initial
value $s^{(i)}_z$, so that the retrieved final value $s^{(f)}_z=u_1 +
u_2 s^{(i)}_z$ (i.e., the value $s^{(f)}_z$ is linearly dependent on
$s^{(i)}_z$ but not proportional to it).  This type of error is, as
far as we know, irrecoverable.  To quantify it, we consider the value
of $s^{(f)}_z$ for the case $s_z^{(i)}=1$, $s_x^{(i)}=s_y^{(i)}=0$,
and the value of $s^{(f)}_z$ for the initial conditions
$s_x^{(i)}=1$, $s_y^{(i)}=s_z^{(i)}=0$. For the latter case, the value
$s^{(f)}_z$ is denoted as $s_0$, to distinguish it from the value
$s^{(f)}_z$ obtained with the initial conditions $s_z^{(i)}=1$,
$s_x^{(i)}=s_y^{(i)}=0$.  $s_z$ and $s_0$ are the remaining two
performance measures used in the paper.
In ideal situation of $\Delta P=0$, $A_k=A$,
the value $s_0$ should be zero, while in a non-ideal situation, $s_0$
quantifies the protocol error.  Figs.~\ref{fig1}(e--h) illustrate this
point, showing the oscillations of $s_z$ and $s_0$ for $\Delta P=0$
(left column, (e) and (g)), and for $\Delta P=0.2$ (right column,(f)
and (h)).  The solid lines correspond to $N=20$, and the dashed lines
correspond to $N=10^4$. Again, during the first oscillation period,
the results obtained with $N=20$ are close to those obtained with
$N=10^4$, so that $N=20$ is sufficient to reliably estimate the
fidelity of the quantum memory protocol.

Now we study performance of the quantum memory protocol at
a quantitative level. The goal is to ensure that consideration of
$N=20$ is sufficient for making conclusions about the realistic case
of $N\sim 10^4$--$10^6$.  For $A_k=A$, using the
Eqs.~\ref{oscills},\ref{wIM}, the evolution of the reduced density
matrix of the electron spin can be computed for the whole two-step
protocol for $N\le 10^8$.  The spin ejection is equivalent to a
measurement of $S^z$ described as a von Neumann's projection, so that
at the ejection time $t=t_e$, the wave function
\begin{equation}
|\psi(t_e)\rangle = |\uparrow\rangle\otimes|\nu_1\rangle +
  |\downarrow\rangle\otimes|\nu_2\rangle
\end{equation}
is transformed into a mixed state with the compound system's density matrix
\begin{equation}
R_e = |\uparrow\rangle\langle\uparrow|\otimes|\nu_1\rangle\langle\nu_1| +
  |\downarrow\rangle\langle\downarrow|\otimes|\nu_2\rangle\langle\nu_2|\ .
\label{proj1}
\end{equation}
At the retrieval step, after injection of another electron
in a dot, the compound system's density matrix becomes
\begin{equation}
R_r = |\downarrow\rangle\langle\downarrow|\otimes(|\nu_1\rangle\langle\nu_1| +
  |\nu_2\rangle\langle\nu_2|)\ .
\label{proj2}
\end{equation}
The system's subsequent evolution can be again calculated using
Eqs.~\ref{oscills}. The resulting analytical expressions are
complicated, but they can be easily estimated numerically for $N\le 10^8$.

\begin{figure}
\includegraphics[width=8cm]{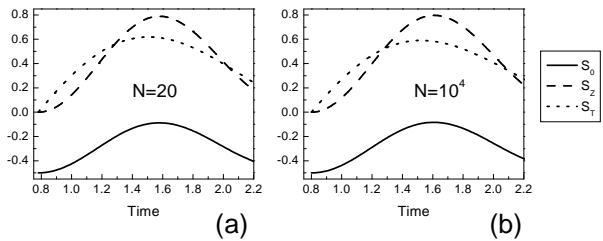}
\caption{Final step of the protocol, retrieval. The values
$s_0$ (solid line), $s_z$ (dashed line) and $s_T$
(dotted line) are shown, obtained for $\Delta P=0.2$ 
with $N=20$ (a) and $N=10^4$ (b).}
\label{fig2}
\end{figure}

Fig.~\ref{fig2} illustrates the final step of the protocol, the
retrieval of the initial state of the electronic spin, for
$\Delta P=0.2$ for N=$20$ (a) and $N=10^4$ (b).  All three parameters
quantifying the performance, $s_T$, $s_z$ and $s_0$, are shown for
times close to the time of the state retrieval (half-period after the
injection of the second electron), and one can see that the curves for
$N=20$ almost exactly replicate the curves for $N=10^4$ (the
difference is $\sim 0.05$).  Moreover, it is important to note that
the maximum performance of the retrieval is achieved at slightly
different times for $s_T$, $s_z$ and $s_0$. Below, we assume that the
information retrieval is performed when $s_z$ reaches the maximum.  In
section~\ref{s:numerics} we evaluate all three performance measures,
$s_z, s_T,$ and $s_0$, while calculation of the fidelity of the
protocol for arbitrary quantum states is considered in
section~\ref{s:fidelity}.

\section{Numerics for $A_k\neq A$:
computations for $N=20$ and analysis \label{s:numerics}}

In a real quantum dot, the electron density is non-uniform, so that
the values of the hyperfine coupling constants $A_k$ differ
substantially across the dot. Formally, the Hamiltonian (\ref{hamt})
can be solved analytically for $A_k\neq A$ by Bethe ansatz
\cite{gaudin76}.  However, this formal solution is impractical,
especially in our problem, where the state of the bath has a complex
form, with a large number of excited states, and where the non-unitary
projection (see Eqs.~\ref{proj1},\ref{proj2}) is an important part of
the protocol. We are not aware of any way of employing the formally
exact solution of Ref.~\onlinecite{gaudin76} (which requires solving a
system of order of $2^N/\sqrt{N}$ nonlinear equations) for practical
quantitative estimates in such complex problems as the one considered
here, even for a small number of the bath spins. For similar reasons,
numerical approaches based on the P-representation of the spin density
matrix \cite{khaled} are also  not applicable. Instead, we use the
numerical approach suggested in
Ref.~\onlinecite{dobrovitski03b,deraedt04}: the direct solution of the
time-dependent Schr\"odinger equation for the compound system (the
electron spin plus the bath spins).  This allows straightforward
numerical modeling of the quantum memory protocol.  The drawback of
such an approach is that the exponential scaling of computation costs
with $N$ limits simulations to $N\sim 20$ spins.  Considering the
above case of $A_k=A$, we have shown that $N=20$ is sufficient to make
our simulations reflect larger $N$ behavior with a precision of order 0.05.

In order to understand whether this remains true for $A_k\neq$ {\em const},
let us consider the case of perfect polarization $P=1$ which
can be studied analytically with precision of $1/N$ for large $N$. 
First, let us consider
the electron initially in the state $|\phi_0\rangle=|\uparrow\rangle$ and
the bath in the state $|\chi_0\rangle=|\downarrow,\downarrow,\dots\rangle$.
By acting with the Hamiltonian (\ref{hamt}) on the state
$|\psi_0\rangle\otimes|\chi_0\rangle$, we find that
\begin{equation}
{\cal H}|\psi_0\rangle = (g_e\mu_B/2) (H_0+H_{\text Ovh})|\psi_0\rangle +
(b/2) |\psi_1\rangle
\end{equation}
where $H_{\text Ovh}=-\sum_k A_k/(2g_e\mu_B)$ is the Overhauser field
acting from the nuclei on the electron spin, $b=\sqrt{\sum_k A_k^2}$, and
$|\psi_1\rangle=|\downarrow\rangle\otimes|\chi_1\rangle=
|\downarrow\rangle\otimes[(1/\sqrt{N})\sum_k S_k^+ |\chi_0\rangle]$.
Analogous result for $|\psi_1\rangle$ can be presented in the form
\begin{eqnarray}
{\cal H}|\psi_1\rangle &=& -(g_e\mu_B/2) (H_0+H_{\text Ovh}+M_3/M_2)|\psi_1\rangle \\
 \nonumber
  &+& (b/2) |\psi_0\rangle + (b/2) |u'\rangle
\end{eqnarray}
where $M_n=\sum_k A_k^n$ is the $n$-th moment of the distribution
of the hyperfine couplings $A_k$ (correspondingly, $b=\sqrt{M_2}$),
and the extra state $|u'\rangle$ is orthogonal to both $|\psi_0\rangle$ and
$|\psi_1\rangle$. By choosing the external field equal to its optimal value
\begin{equation}
H_0 = -H_{\text Ovh} - M_3/(2 M_2 g_e\mu_B)
\label{h0opt}
\end{equation}
and subtracting a constant from the Hamiltonian, we have
\begin{eqnarray}
\bar{\cal H}|\psi_0\rangle &=& (b/2) |\psi_1\rangle \\ \nonumber
\bar{\cal H}|\psi_1\rangle &=& (b/2) |\psi_0\rangle + (b/2) |u'\rangle
\end{eqnarray}
where $\bar{\cal H} = {\cal H}+M_3/(2 M_2)$.

This system is ``practically'' closed, since the norm of the
extra state $|u'\rangle$ is small for large $N$,
$||u'||^2=[(M_4/M_2^2)-(M^2_3/M^3_2)]\sim 1/N$ for a ``non-pathological''
distribution of $A_k$. 
As a result, the calculations for both the encoding and the retrieval 
steps can be done explicitly with the precision of $1/N^2$. Introducing new
basis states $|\psi_+\rangle = (1/\sqrt{2})[|\psi_1\rangle+|\psi_0\rangle]$, 
$|\psi_-\rangle=(1/\sqrt{2})[|\psi_1\rangle-|\psi_0\rangle]$, 
and $|u'/||u'||\rangle$, the Hamiltonian $\bar{\cal H}$ can be represented 
by the matrix
\begin{equation}
\bar{\cal H}=(b/2)\left(
\begin{array}{ccc}
1 & 0 & \delta \\ 
0 & -1 & -\delta \\ 
\delta & -\delta & 0
\end{array}
\right)
\end{equation}
where $\delta=||u'||/\sqrt{2}$. The unitary evolution during the
encoding and the retrieval steps is described by the evolution
operator $\exp{(-i{\bar{\cal H}}t)}$ which can be easily calculated,
as well as the system's evolution during the projection step.
We also note that the optimal times for encoding and for retrieval
differ slightly (by the terms of order of $\delta^2$) from $\pi/(2b)$,
but this difference can be neglected since it leads to the corrections 
of order of $\delta^3$ during the encoding/retrieval steps.

By performing the straightforward calculations, one can arrive at the
final answers for the performance parameters $s_T$, $s_0$, and $s_z$:
\begin{equation}
s_0 = -8\delta^2,\quad s_T=1-4\delta^2,\quad s_z=1-16\delta^2
\label{analytfid}
\end{equation}
The deviation from ideal values is of order of $\delta^2\sim 1/N$,
which is about 5\% for our simulations with $N=20$ (below, we will
demonstrate the agreement in more detail). 

\begin{figure}
\includegraphics[width=8cm]{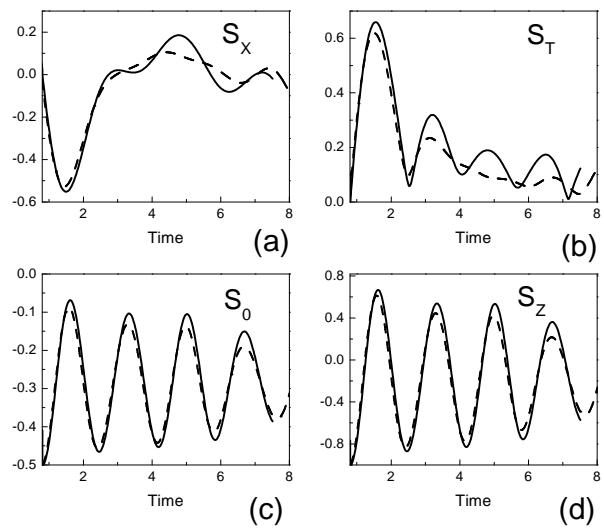}
\caption{Final step of the protocol, retrieval. The values
$s_x$ (a), $s_T$ (b), $s_0$ (c), and $s_z$ (d) are shown, 
obtained for $\Delta P=0.2$ 
with $N=20$. The solid curves represent the results obtained
with the bath's initial conditions obtained by application
of the operator $\exp{(-\gamma I^z)}$ to the random state,
the dashed curves correspond to the analytical solution.}
\label{fig3}
\end{figure}

In our simulations, two types of 
initial conditions are considered: (i) the electronic spin is in the state 
$s_z^{(i)}=1$, $s_x^{(i)}=s_y^{(i)}=0$, for evaluation of the $s_z$
performance measure,
and (ii) the electronic spin is in the state
$s_x^{(i)}=1$, $s_z^{(i)}=s_y^{(i)}=0$, for evaluation of the $s_T$
and $s_0$ performance measures. Then we can  use
the values $s_z$, $s_T$, and $s_0$ to determine the fidelity
of the protocol (see section~\ref{s:fidelity}). In order to simulate the initial state of the 
polarized bath, we take the bath's state as a random 
superposition $|r\rangle$ of all possible basis states, 
apply the operator $\exp{(-\gamma I^z)}$ to the state $|r\rangle$
(such an operator is easily implemented using the Chebyshev's
polynomials expansion described in detail in Ref.~\onlinecite{dobrovitski03b,deraedt04}),
and normalize the resulting state.
Such a state corresponds to the bath's density matrix 
$\rho=(1/Z)\exp{(-\gamma I^z)}$.

\begin{figure}
\includegraphics[width=8cm]{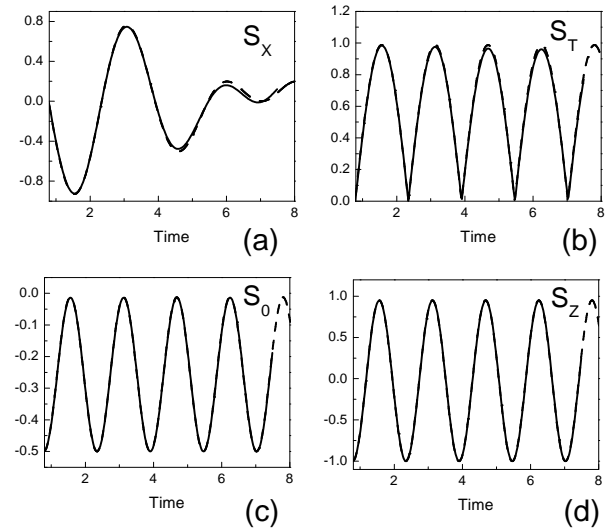}
\caption{Final step of the protocol, retrieval. The values
$s_x$ (a), $s_T$ (b), $s_0$ (c), and $s_z$ (d) are shown, 
obtained for $\Delta P=0$ 
with $N=20$. The solid curves represent the results obtained
with the bath's initial conditions obtained by application
of the operator $\exp{(-\gamma I^z)}$ to the random state,
the dashed curves correspond to the analytical solution.}
\label{fig4}
\end{figure}

This construction of the bath's initial conditions can lead to statistical
errors if the number of relevant states is not large enough. 
To ensure that the constructed
initial condition is valid for $N=20$, we did the simulations with
$A_k=A$, and compared them with the analytical results given
in the previous section. The comparison is presented in 
Figs.~\ref{fig3},\ref{fig4} for $\Delta P=0.2$ (Fig.~\ref{fig3})
and $\Delta P=0$ (Fig.~\ref{fig4}), for $s_z$, $s_0$, $s_T$, and
also for $s_x$ itself. Even in worst case of $\Delta P=0.2$,
the simulation results remain close to the analytical ones,
and at times close to the information retrieval time,
the difference does not exceed $\sim 0.05$. For smaller
$\Delta P$, the difference is smaller, and almost disappears
for $\Delta P=0$, see Fig.~\ref{fig4}.

\begin{figure}
\includegraphics[width=8cm]{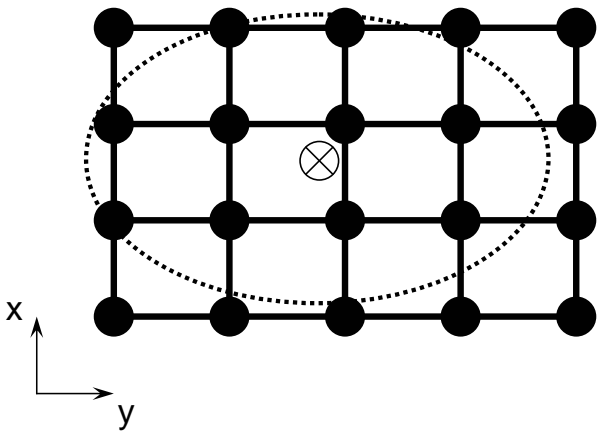}
\caption{The arrangement of the lattice of the bath spins
and the electron density.
The bath spins sites are marked by the solid circles,
and the position of the maximum of the electron density
is marked as $\otimes$; the dashed line schematically shows 
the contour line where the electron density is equal to
$\exp{(-1/2)}$ of its maximum value.}
\label{fig5}
\end{figure}

For the simulations, we consider 
$N=20$ bath spins which are placed at the sites of a $4\times 5$
piece of a rectangular lattice with the lattice constants $a_y$
and $a_x$. We assume that the electrostatic potential inside the dot
can be approximated as a 2-D harmonic well, so that
the electron density in the dot is a 2-D Gaussian with
the half-widths $w_x=(3/2) a_x$ and $w_y=2 a_y$ along the 
$x$- and $y$-axes. We set the maximum
of the electron density is slightly displaced with respect 
to the center of the lattice by $l_y=0.2$, $l_x=0.1$ to prevent
artificial symmetry effects.
This arrangement is schematically shown on Fig.~\ref{fig5},
where the bath spins' sites are marked by the solid circles,
and the position of the maximum of the electron density
is marked as $\otimes$; the dashed line schematically shows 
the contour line where the electron density is equal to
$\exp{(-1/2)}$ of its maximum value. The values of $A_k$
were taken proportional to the electron density with
the proportionality factor 1, and were spread from 0.96 to
0.31. 

\begin{figure}
\includegraphics[width=8cm]{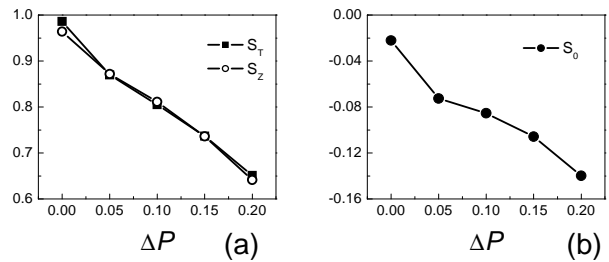}
\caption{Performance of the memory protocol as a function of
$\Delta P$ for $w_x=(3/2) a_x$ and $w_y=2 a_y$. 
(a): the values of $s_T$ (solid squares) and $s_z$ 
(open circles); (b): the values of $s_0$ (solid circles).
The values of $s_T$ and $s_z$ are very close to each
other, and the corresponding curves almost coincide on
the graph.}
\label{fig6}
\end{figure}

The external field $H_0$ in our simulations was taken 
equal to its optimal value, which compensates the Overhauser field 
$H_{\text Ovh}=-P \sum_k A_k/(2g_e\mu_B)$ plus
a small correction,
$H_0 = -H_{\text Ovh} - M_3/(2 M_2 g_e\mu_B)$
(see Eq.~\ref{h0opt}).
The memory protocol was simulated as follows:
\begin{enumerate}
\item The bath's initial polarized state was obtained by acting
with the operator $\exp{(-\gamma I^z)}$ to the random state;
the initial state of the electron spin was $|\uparrow\rangle$.
\item The electron spin is allowed to interact with the bath,
and the time when $s_z$ reaches its minimum is assumed to
be the ejection time $t_e$.
\item The ejection of the first electron is described according
to Eq.~\ref{proj1}.
\item The second electron is injected in the state $|\downarrow\rangle$,
so the compound system's density matrix has the form of Eq.~\ref{proj2}.
\item The second electron is allowed to interact with the bath,
and the time when $s_z$ reaches its maximum is assumed to
be the retrieval time $t_r$.
\end{enumerate}
The fidelity of the protocol is indirectly determined by the values of
$s_T$, $s_z$, and $s_0$ at time $t_r$: for ideal memory,
$s_T(t_r)$ and $s_z(t_r)$ would be one, and $s_0(t_r)$ would be zero.

The results for different values of $\Delta P$ are presented in
Fig.~\ref{fig6}. 
Both
$s_z$ and $s_T$ are very close to each other,
and both decrease approximately
linearly with decreasing polarization. Similarly, $s_0$ deviates from
zero linearly with decreasing polarization.

\begin{figure}
\includegraphics[width=8cm]{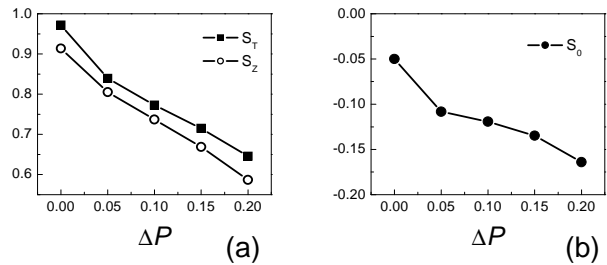}
\caption{Performance of the memory protocol as a function of
$\Delta P$ for $w_x=3 a_x/(2\sqrt{2})$ and $w_y=\sqrt{2} a_y$. 
(a): the values of $s_T$ (solid squares) and $s_z$ 
(open circles); (b): the values of $s_0$ (solid circles).
In spite of noticeably larger spread of the couplings $A_k$,
retrieval values decrease only by 5--7\% (depending on $\Delta P$)
in comparison with Fig.~\ref{fig6}.}
\label{fig7}
\end{figure}

For comparison, in Fig.~\ref{fig7}, we present the results
for the case of $w_x$ and $w_y$ decreased by a factor of $1/\sqrt{2}$
with respect to Fig.~\ref{fig6} (so that the
dashed line of Fig.~\ref{fig5} corresponds now to the contour 
of the electron density equal to
${\text e}^{-1}$ of the maximum value). This corresponds to
a noticeably larger spread of $A_k$, from 0.92 to 0.09. 
One can see that the perfromance of the protocol is determined by the 
value of $s_z$. In spite of considerably larger spread
of $A_k$, the change in performance measures $s_z,s_0, s_T$  is small, from 
5\%--7\% depending on the bath polarization.
This is expected: the smaller values of $A_k$
give smaller contribution to the evolution, and, correspondingly,
do not produce large errors in the course of the system's evolution.

\begin{figure}
\includegraphics[width=8cm]{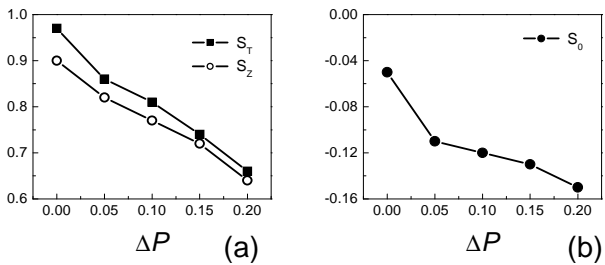}
\caption{Performance of the memory protocol as a function of
$\Delta P$ for the unoptimized external field, for
$w_x=(3/2) a_x$ and $w_y=2 a_y$. 
(a): the values of $s_T$ (solid squares) and $s_z$ 
(open circles); (b): the values of $s_0$ (solid circles).
The maximum drop in performance in comparison with Fig.~\ref{fig6}
is only about $6\%$, and is noticeable only for small $\Delta P$.}
\label{fig8}
\end{figure}

We note here that the quantum memory protocol described in the
introduction implies fine adjustment of the external field. 
The optimal value of $H_0$ includes the small correction
$M_3/(2 M_2)\sim 1/N$. While the Overhauser field
can be measured rather precisely, the small correction $M_3/(2M_2)$
could be hard to determine in practice. However, this is a $1/N$
correction.  According to our
calculations, omission of this correction does not affect much 
the performance of the scheme. Fig.~\ref{fig8} shows the results
obtained for the same parameters as Fig.~\ref{fig6}, except for
the unoptimized external field (i.e., $H_0=-H_{\text Ovh}$, without
the correction $M_3/(2M_2)$). The maximum drop in performance is only
6\% (for $\Delta P=0$), and goes to almost zero with increasing $\Delta P$.

Thus, our results show that the quantum memory protocol remains
feasible for realistic situations, where the polarization of the
nuclear spins is incomplete, and the values of hyperfine couplings
$A_k$ are spread over a large interval. Also, the external field does
not need to be adjusted with precision of $1/N\sim 10^{-4}$ for the
quantum memory protocol to work.  We believe that our results for
$N=20$ provide adequate estimates of the performance of the protocol
for realistic dots with $N=10^4$.  In order to understand the
corrections associated with extending from $N=20$ to larger $N$, let
us compare the analytical estimates given in Eq.~(\ref{analytfid})
with the numerical results.  For the case shown in Fig.~\ref{fig6},
the analytics gives $s_T=1.0-0.008$, $s_0=-0.016$, and $s_z=1.0-0.03$,
while the numerical simulations give $s_T=1.0-0.014$, $s_0=-0.022$,
and $s_z=1.0-0.036$. Similarly, for the case shown in Fig.~\ref{fig7},
the analytics gives $s_T=1.0-0.021$, $s_0=-0.042$, and
$s_z=1.0-0.084$, and the numerical results are $s_T=1.0-0.028$,
$s_0=-0.05$, and $s_z=1.0-0.082$. The agreement between simulations
and analytical results is reasonable, and we can consider our
simulations to have a precision of 5--8\%.

\section{Minimal fidelity of the memory protocol \label{s:fidelity}}

We now show how the values of $s_0$, $s_T$ and $s_z$ calculated
above determine the fidelity of the memory protocol.
The electron initially is in the state 
$|\phi_0\rangle = \alpha|\downarrow\rangle + \beta|\uparrow\rangle$,
so that the electron's reduced density matrix initially is
$\tau_0=|\phi_0\rangle\langle\phi_0|$, and the density matrix of
the compound system is $R_0=\tau_0\otimes\rho$, where $\rho$ is the
initial density matrix of the bath.
At the end of the retrieval stage, the final state of the compound system
is described by the density matrix $R_f$, so that the electron's
reduced density matrix is $\tau_f={\mathrm{Tr}}_B R_f$, where
${\mathrm{Tr}}_B$ is the trace over the bath spins.
Ideally, we want $\tau_f=\tau_0$, but in reality the initial and
the final density matrices of the electron are different.
The fidelity for the protocol for the density matrix $\tau_0$ is 
$F_{\tau_0} = {\rm Tr}[\tau_0 \tau_f] = \langle\phi_0 | \tau_f |\phi_0\rangle$.  
Accordingly, the minimal performance of the memory protocol (the fidelity, $F$) 
is defined in a standard way \cite{qcqi}, as
\begin{equation}
F ={\rm Min}_{\phi_0} \left[ \langle\phi_0|\tau_f|\phi_0\rangle \right],
\end{equation}
which is minimized over all possible initial states $|\phi_0\rangle$.

To calculate the fidelity, we use the ideas of the quantum process 
tomography \cite{qcqi}. First, let us analyze in detail the dynamics of
the density matrix. At the encoding stage,
the evolution of the compound system is described
by the unitary operator $U_1=\exp{[-i{\cal H}t_e]}$, so by the
end of the encoding stage, the state of the compound system
is $R_1=U_1 R_0 U_1^\dag = U_1 (\tau_0\otimes\rho) U_1^\dag$.
During the ejection of the first electron and injection of
another one, the evolution is not unitary (see Eqs.~\ref{proj1},\ref{proj2});
after injection of the second electron, the state of the
compound system is:
\begin{equation}
R_e=P_{\downarrow}R_1P_{\downarrow} + S^{-} R_1 S^{+},
\end{equation}
where $P_{\downarrow}=|\downarrow\rangle\langle\downarrow|$ is 
the projection operator on the state $|\downarrow\rangle$ of
the electron spin.
At the retrieval stage, the evolution is again unitary, described
by the operator $U_2=\exp{[-i{\cal H}(t_r-t_e)]}$ so the
final state of the compound system is
\begin{eqnarray}
\nonumber
R_f&=&U_2 P_{\downarrow} U_1 (\tau_0\otimes\rho) 
  U_1^{\dag} P_{\downarrow} U_2^{\dag} \\
  &+& U_2 S^{-} U_1 (\tau_0\otimes\rho) U_1^{\dag} S^{+} U_2^{\dag}.
\label{totevol}
\end{eqnarray}

In order to analyze the evolution of the electron's reduced density matrix,
we consider the evolution superoperator $\cal L$ for the electron spin
\begin{equation}
{\cal L}[\tau_0] = \tau_f = {\mathrm{Tr}}_B R_f.
\label{superop}
\end{equation}
The superoperator $\cal L$ is linear. By representing the initial 
electron's density matrix in the form 
$\tau_0=(1/2){\mathbf 1} + b_x S^x + b_y S^y + b_z S^z$,
where ${\mathbf 1}$ is the identity matrix $2\times 2$,
we obtain for the final density matrix $\tau_f$:
\begin{equation}
\tau_f={\cal L}[\tau_0]=\tau_f^0 + b_x\tau_f^x + b_y\tau_f^y + b_z\tau_f^z
\label{taufin}
\end{equation}
where
\begin{eqnarray}
\nonumber
\tau_f^0&=&{\cal L}[(1/2){\mathbf 1}],\quad \tau_f^x={\cal L}[S^x],\\
\tau_f^y&=&{\cal L}[S^y],\quad \tau_f^z={\cal L}[S^z].
\end{eqnarray}
Each of the matrices $\tau_f^0,\dots \tau_f^z$ can be expanded in
the basis ${\mathbf 1}$, $S^x$, $S^y$, and $S^z$, so we just 
need to determine the expansion coefficients.
This problem is considerably simplified by the fact that the superoperator
$\cal L$ is invariant with respect to the rotations of the compound system
in the $x$--$y$ plane, i.e.\ with respect to the operators
$\exp{[ia J^z]}$, where $J^z=S^z+\sum_k I_k^z$ and $a$ is the rotation angle.
For example, let us   
apply the rotation by the angle $\pi$ in the $x$--$y$ plane
(the operator $\exp{[i\pi J^z]}$) to the matrix $\tau_f^0$. 
By doing that, we obtain that
\begin{equation}
{\mathrm{Tr}}(\tau_f^0 S^x) = 
  -{\mathrm{Tr}}(\exp{[-i\pi J^z]}\tau_f^0\exp{[i\pi J^z]} S^x).
\end{equation}
On the other hand, $\exp{[-i\pi J^z]}\tau_f^0\exp{[i\pi J^z]}=\tau_f^0$,
so that ${\mathrm{Tr}}(\tau_f^0 S^x)=0$. 
By applying similar symmetry arguments, we derive the following results:
\begin{eqnarray}
\nonumber
\tau_f^0&=&(1/2){\mathbf 1} + S^z u_0\\
\nonumber
\tau_f^z&=&S^z w_0\\
\nonumber
\tau_f^x&=&S^x v_0\cos{\xi} + S^y v_0\sin{\xi},\\
\nonumber
\tau_f^y&=&S^y v_0\cos{\xi} - S^x v_0\sin{\xi},
\label{taus}
\end{eqnarray}
where the last equation is obtained by applying the operator
$\exp{[i(\pi/2)J^z]}$ (rotation by $\pi/2$ in the $x$--$y$ plane)
to the matrix $\tau_f^x$, and noticing that 
$\exp{[-i(\pi/2)J^z]} S^x \exp{[i(\pi/2)J^z]} = S^y$.
In Eqs.~\ref{taus}, $u_0$, $w_0$, $v_0$, and $\xi$
are numerical parameters characterizing the fidelity, which are
related in a simple manner to the parameters
$s_0$, $s_T$, and $s_z$ obtained above from numerical simulations. 
We obtain $s_z$ by using the
initial electron's density matrix $\tau_0=(1/2){\mathbf 1} + S^z$,
so that $s_z=u_0+w_0$. Similarly, $s_0$ and $s_T$ are obtained
from the initial electron's density matrix 
$\tau_0=(1/2){\mathbf 1} + S^x$, so that $s_0=u_0$ and
$s_T=v_0$.
We note that the action of the superoperator $\cal L$
transforming $S^x$ and $S^y$ into $\tau_f^x$ and $\tau_f^y$
involves the rotation in the $x$--$y$ plane by the angle $\xi$.
As we explained above, this rotation can be, in principle, corrected,
and we do not take it into account. Correspondingly, we neglect 
the reduction of fidelity caused by non-zero value of $\xi$,
by putting $\xi=0$.

\begin{figure}
\includegraphics[width=8cm]{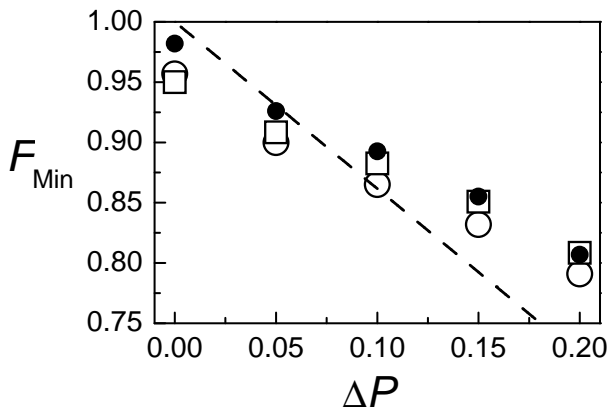}
\caption{Minimal fidelity $F_{\rm Min}$ of the memory protocol as a function of
$\Delta P$. Black circles: for $w_x=(3/2) a_x$ and $w_y=2 a_y$
(same parameters as in Fig.~\ref{fig6}).
Open circles: for $w_x=3 a_x/(2\sqrt{2})$ and $w_y=\sqrt{2} a_y$
(same parameters as in Fig.~\ref{fig7}).
Open squares: unoptimized external field, for
$w_x=(3/2) a_x$ and $w_y=2 a_y$ (same parameters as in Fig.~\ref{fig8}).
The dashed line corresponds to the analytical solution (\ref{FminSmallD})
for small $\Delta P$.
}
\label{fig9}
\end{figure}

By combining Eqs.~\ref{taufin} and \ref{taus}, the fidelity $F_{\tau_0}$
for a given initial density matrix 
$\tau_0=(1/2){\mathbf 1} + b_x S^x + b_y S^y + b_z S^z$ can be
expressed as
\begin{equation}
F_{\tau_0}=(1/2)\left[ 1 + v_0 + b_z u_0 + b_z^2 (w_0-v_0)\right],
\label{fid}
\end{equation}
where we used the fact that for initial pure state of the electron,
$b_x^2+b_y^2=1-b_z^2$, and we need to minimize the fidelity
(\ref{fid}) over all possible values of $b_z$, i.e.\ over all $b_z\in
[-1,1]$.  The minimum value of this expression is achieved either at
the extremum point $b_{z0}=-u_0/[2(w_0-v_0)]$, or at the ends of the
interval.  Correspondingly, the minimal fidelity $F_{\rm Min}$ of the
protocol is equal to the smallest of the three quantities
\begin{eqnarray}
\nonumber
F_1 &=& \frac{1+s_z}{2},\\
\nonumber
F_2 &=& \frac{1+s_z-2s_0}{2},\\
F_3 &=& \frac{1}{2}\bigl(1+s_T-s_0^2/[4(s_z-s_0-s_T)]\bigr)\ .
\end{eqnarray}
With this fidelity measure, our results described above and 
depicted in Figs.~\ref{fig6}, \ref{fig7}, and \ref{fig8}, 
can be presented as a single graph, Fig.~\ref{fig9}. One can see
that the minimum fidelity is $F_{\rm Min} \sim 75$\% for $P=80$\%, while as $P
\rightarrow 100\%$, $F_{\rm Min} \rightarrow 96\%$. The fact that
the latter value is different from 100\% is due to the
modest number of bath spin $N$. Indeed, as seen from Eq.~\ref{analytfid},
for large $N$, the value of $F_{\rm Min}$ for completely polarized
bath differs from unity by the terms of order of $1/N$, and therefore
goes to zero for large $N$.

Using the analytical approach described in Sec.~\ref{QualitAnalyt}, 
we can  estimate
the minimum fidelity for small $\Delta P$ in the approximation
of all $A_k=A=$ {\em const}. Indeed, the most significant contribution
to $s_z$, $s_0$, and $s_T$ is given by the states with $I=-M$
(note that in our consideration $M<0$); the contribution of the states 
with $I=-M+n$ is of order  $(\Delta P)^n$ (where $n$ is a positive
integer). Thus, to calculate the values $s_z$, $s_0$, and $s_T$
up to linear terms in $\Delta P$, we take into account only
two states, $I=-M$ and $I=-M+1$, which have relative statistical
weight $1-\Delta P/2$ and $\Delta P/2$, correspondingly,  
and perform averaging over all $M$ according to Eq.~\ref{wIM}.
The calculations give the following results:
\begin{eqnarray}
\nonumber
s_z&=& 1 - \frac{\Delta P}{2}(5+\cos^4{\gamma_0}+\sin^4{\gamma_0})
 + O((\Delta P)^{3/2})\\ \nonumber
s_0&=& -\frac{\Delta P}{4}(5-\cos^4{\gamma_0}-\sin^4{\gamma_0})
 + O((\Delta P)^{3/2})\\ 
s_T&=& 1 - \frac{\Delta P}{2}(2-\cos{\gamma_0})
 + O((\Delta P)^{3/2})
\end{eqnarray}
where $\gamma_0=\pi/\sqrt{2}$, and from these values we obtain 
\begin{equation}
F_{\rm Min}=1-1.38386 \Delta P + O((\Delta P)^{3/2})\ .
\label{FminSmallD}
\end{equation}
This line is shown in Fig.~\ref{fig9} as a dashed line. 
The difference between this analytical result and the numerical data
comes either from the assumption $A_k=A=$ {\em const} used in the derivation
above, and from the $(\Delta P)^{3/2}$ corrections, which can be important
for $\Delta P ~ 0.1$.

Visible deviation of the fidelity from unity at 100\% polarization results
from finite size of the sample used in our simulations. Indeed, the
analytical calculations of Ref.~\onlinecite{taylor03} and Section 
\ref{s:numerics} of
this paper suggests that at 100\% polarization the error arises 
from the inhomogeneous hyperfine coupling, and hence scales as $1/\sqrt{N}$.
While it corresponds to few percent for  the simulations presented in
Fig.~\ref{fig9} ($N=20$), for realistic dots ($N=10^6$), 
it is well below $10^{-3}$.

\section{Conclusions}

We study numerically the long-lived memory for electronic spin
in a quantum dot taking into account the errors inherent in the
protocol for a realistic system: the incomplete polarization of the bath
of nuclear spins and the spread in the hyperfine interactions
between the electron and the nuclei in the quantum dot.
Although the imperfections deteriorate the fidelity of the
quantum information retrieval, 
the scheme remains feasible for the spin bath 
polarizations exceeding 80\%, with a minimum fidelity 
of $F\sim 80$\%. Fidelity is determined mostly
by the retrieved value of $s_z$, and increases linearly
with increasing polarization of the bath.

Since direct simulation of a real system with $\sim 10^4$ nuclear
spins is currently impossible, we consider the system of $N=20$ spins
coupled to the electron spin. Such a step requires 
justification, and accordingly a major part of our work is devoted to 
understanding the reliability of our simulations. By comparing 
different aspects of the numerical calculations with the analytical
results, we show that our results give adequate estimates for 
realistic situations, with a precision of 5--8\%.  

The errors considered in this work may be reduced by changes to the
storage protocol or altered by considering of more realistic initial
conditions.  For example, it is expected that at high polarizations,
created by electron spin-nuclear spin transfer techniques (e.g.,
Ref.~\onlinecite{imamoglu03}), the nuclear spins may be in states with
favorable symmetry properties~\cite{taylor03b} allowing for reduced
error due to polarization.  Similarly, estimating the current value of
the nuclear spin polarization reduces the error associated with the
uncertainty in polarization~\cite{giedke05,klauser05}.  More detailed
work will be necessary to consider the effects of these complexities.

\begin{acknowledgments}
V.~V.~D. thanks Harvard University for hospitality.
This work was supported in
part by the National Security Agency (NSA) and Advanced
Research and Development Activity (ARDA) under Army Research
Office (ARO) contracts DAAD 19-03-1-0132.
This work was partially
carried out at the Ames Laboratory, which is operated
for the U. S. Department of Energy by Iowa State
University under Contract No. W-7405-82 and was supported
by the Director of the Office of Science, Office
of Basic Energy Research of the U. S. Department of
Energy.
\end{acknowledgments}


\begin{thebibliography}{19}
\expandafter\ifx\csname natexlab\endcsname\relax\def\natexlab#1{#1}\fi
\expandafter\ifx\csname bibnamefont\endcsname\relax
  \def\bibnamefont#1{#1}\fi
\expandafter\ifx\csname bibfnamefont\endcsname\relax
  \def\bibfnamefont#1{#1}\fi
\expandafter\ifx\csname citenamefont\endcsname\relax
  \def\citenamefont#1{#1}\fi
\expandafter\ifx\csname url\endcsname\relax
  \def\url#1{\texttt{#1}}\fi
\expandafter\ifx\csname urlprefix\endcsname\relax\def\urlprefix{URL }\fi
\providecommand{\bibinfo}[2]{#2}
\providecommand{\eprint}[2][]{\url{#2}}

\bibitem[{\citenamefont{Loss and DiVincenzo}(1998)}]{loss98}
\bibinfo{author}{\bibfnamefont{D.}~\bibnamefont{Loss}} \bibnamefont{and}
  \bibinfo{author}{\bibfnamefont{D.}~\bibnamefont{DiVincenzo}},
  \bibinfo{journal}{Phys. Rev. A} \textbf{\bibinfo{volume}{57}},
  \bibinfo{pages}{120} (\bibinfo{year}{1998}).

\bibitem[{\citenamefont{Elzerman et~al.}(2004)\citenamefont{Elzerman, Hanson,
  van Beveren, Witkamp, Vandersypen, and Kouwenhoven}}]{elzerman04}
\bibinfo{author}{\bibfnamefont{J.~M.} \bibnamefont{Elzerman}}
 {\em et al.}
\bibinfo{journal}{Nature}
  \textbf{\bibinfo{volume}{430}}, \bibinfo{pages}{431} (\bibinfo{year}{2004}).

\bibitem[{\citenamefont{Johnson et~al.}(2005)\citenamefont{Johnson, Petta,
  Taylor, Lukin, Marcus, Hanson, and Gossard}}]{johnson05}
\bibinfo{author}{\bibfnamefont{A.~C.} \bibnamefont{Johnson}}
 {\em et al.}
\bibinfo{journal}{Nature}
  \textbf{\bibinfo{volume}{435}}, \bibinfo{pages}{925} (\bibinfo{year}{2005}).

\bibitem[{\citenamefont{Petta et~al.}(2005)\citenamefont{Petta, Johnson,
  Taylor, Laird, Yacoby, Lukin, and Marcus}}]{petta05}
\bibinfo{author}{\bibfnamefont{J.}~\bibnamefont{Petta}}
 {\em et al.}
  \bibinfo{journal}{Science}, \textbf{309}, 2180   (\bibinfo{year}{2005}).

\bibitem[{\citenamefont{Imamoglu et~al.}(1999)\citenamefont{Imamoglu,
  Awschalom, Burkard, DiVincenzo, Loss, Sherwin, and Small}}]{imamoglu99}
\bibinfo{author}{\bibfnamefont{A.}~\bibnamefont{Imamoglu}}
 {\em et al.}
  \bibinfo{journal}{Phys. Rev. Lett.} \textbf{\bibinfo{volume}{83}},
  \bibinfo{pages}{4204} (\bibinfo{year}{1999}).

\bibitem[{\citenamefont{Bracker et~al.}(2005)\citenamefont{Bracker, Stinaff,
  Gammon, Ware, Tischler, shabaev, Efros, Park, Gershoni, Korenev
  et~al.}}]{bracker04}
\bibinfo{author}{\bibfnamefont{A.~S.} \bibnamefont{Bracker}}
  {\em et~al.}, \bibinfo{journal}{Phys. Rev. Lett.}
  \textbf{\bibinfo{volume}{94}}, \bibinfo{pages}{047402}
  (\bibinfo{year}{2005}).

\bibitem[{\citenamefont{Awschalom et~al.}(2002)\citenamefont{Awschalom,
  Samarth, and Loss}}]{awschalom01}
\bibinfo{editor}{\bibfnamefont{D.~D.} \bibnamefont{Awschalom}},
  \bibinfo{editor}{\bibfnamefont{N.}~\bibnamefont{Samarth}}, \bibnamefont{and}
  \bibinfo{editor}{\bibfnamefont{D.}~\bibnamefont{Loss}}, eds.,
  \emph{\bibinfo{title}{Semiconductor Spintronics and Quantum Computation}}
  (\bibinfo{publisher}{Springer-Verlag}, \bibinfo{address}{Berlin},
  \bibinfo{year}{2002}).

\bibitem[{\citenamefont{Datta and Das}(1990)}]{datta90}
\bibinfo{author}{\bibfnamefont{S.}~\bibnamefont{Datta}} \bibnamefont{and}
  \bibinfo{author}{\bibfnamefont{B.}~\bibnamefont{Das}},
  \bibinfo{journal}{Appl. Phys. Lett.} \textbf{\bibinfo{volume}{56}},
  \bibinfo{pages}{665} (\bibinfo{year}{1990}).

\bibitem{prokstamp} N. V. Prokof'ev and P. C. E. Stamp, Rep. Prog. Phys. 
 {\bf 63}, 669 (2000).
\bibitem{dkdh} V. V. Dobrovitski, H. A. De Raedt, M. I. Katsnelson, and B. N. Harmon,
 Phys. Rev. Lett. {\bf 90}, 210401 (2003).

\bibitem[{\citenamefont{Taylor et~al.}(2003{\natexlab{a}})\citenamefont{Taylor,
  Marcus, and Lukin}}]{taylor03}
\bibinfo{author}{\bibfnamefont{J.~M.} \bibnamefont{Taylor}},
  \bibinfo{author}{\bibfnamefont{C.~M.} \bibnamefont{Marcus}},
  \bibnamefont{and} \bibinfo{author}{\bibfnamefont{M.~D.} \bibnamefont{Lukin}},
  \bibinfo{journal}{Phys. Rev. Lett.} \textbf{\bibinfo{volume}{90}},
  \bibinfo{pages}{206803} (\bibinfo{year}{2003}{\natexlab{a}}).

\bibitem[{\citenamefont{Taylor et~al.}(2004)\citenamefont{Taylor, Giedke,
  Christ, Paredes, Cirac, Zoller, Lukin, and Imamoglu}}]{taylor04}
\bibinfo{author}{\bibfnamefont{J.~M.} \bibnamefont{Taylor}}
{\em et al.}
  \bibinfo{journal}{e-print: cond-mat/0407640}  (\bibinfo{year}{2004}).

\bibitem[{\citenamefont{Deng and Hu}(2004)}]{deng04}
\bibinfo{author}{\bibfnamefont{C.}~\bibnamefont{Deng}} \bibnamefont{and}
  \bibinfo{author}{\bibfnamefont{X.}~\bibnamefont{Hu}},
  \bibinfo{journal}{e-print: cond-mat/0406478}  (\bibinfo{year}{2004}).

\bibitem[{\citenamefont{Taylor et~al.}(2003{\natexlab{b}})\citenamefont{Taylor,
  Imamoglu, and Lukin}}]{taylor03b}
\bibinfo{author}{\bibfnamefont{J.~M.} \bibnamefont{Taylor}},
  \bibinfo{author}{\bibfnamefont{A.}~\bibnamefont{Imamoglu}}, \bibnamefont{and}
  \bibinfo{author}{\bibfnamefont{M.~D.} \bibnamefont{Lukin}},
  \bibinfo{journal}{Phys. Rev. Lett.} \textbf{\bibinfo{volume}{91}},
  \bibinfo{pages}{246802} (\bibinfo{year}{2003}{\natexlab{b}}).

\bibitem[{\citenamefont{Giedke et~al.}(2005)\citenamefont{Giedke, D'Alessandro,
  Lukin, and Imamoglu}}]{giedke05}
\bibinfo{author}{\bibfnamefont{G.}~\bibnamefont{Giedke}},
J.~M.~Taylor, 
  \bibinfo{author}{\bibfnamefont{D.}~\bibnamefont{D'Alessandro}},
  \bibinfo{author}{\bibfnamefont{M.~D.} \bibnamefont{Lukin}},
  \bibnamefont{and} \bibinfo{author}{\bibfnamefont{A.}~\bibnamefont{Imamoglu}},
  \bibinfo{journal}{e-print: quant-ph/0508144}  (\bibinfo{year}{2005}).

\bibitem[{\citenamefont{Raedt and Dobrovitski}(2004)}]{deraedt04}
H. De Raedt and V. V. Dobrovitski, ``Decoherence
  in Quantum Spin Systems'', in {\it Computer
  Simulation Studies in Condensed-Matter Physics XVI\/}, D. P. Landau,
  S. P. Lewis, and H.-B. Sch\"uttler (eds.) (Springer Verlag, Berlin,
  Heidelberg, New York, 2004).

\bibitem[{\citenamefont{Dobrovitski and Raedt}(2003)}]{dobrovitski03b}
\bibinfo{author}{\bibfnamefont{V.~V.} \bibnamefont{Dobrovitski}}
  \bibnamefont{and} \bibinfo{author}{\bibfnamefont{H.~A.~De}
  \bibnamefont{Raedt}}, \bibinfo{journal}{Phys. Rev. E}
  \textbf{\bibinfo{volume}{67}}, \bibinfo{pages}{056702}
  (\bibinfo{year}{2003}).

\bibitem[{\citenamefont{Paget et~al.}(1977)\citenamefont{Paget, Lampel,
  Sapoval, and Safarov}}]{paget}
\bibinfo{author}{\bibfnamefont{D.}~\bibnamefont{Paget}},
  \bibinfo{author}{\bibfnamefont{G.}~\bibnamefont{Lampel}},
  \bibinfo{author}{\bibfnamefont{B.}~\bibnamefont{Sapoval}}, \bibnamefont{and}
  \bibinfo{author}{\bibfnamefont{V.}~\bibnamefont{Safarov}},
  \bibinfo{journal}{Phys. Rev. B} \textbf{\bibinfo{volume}{15}},
  \bibinfo{pages}{5780} (\bibinfo{year}{1977}).

\bibitem[{\citenamefont{Waugh et~al.}(1968)\citenamefont{Waugh, Huber, and
  Haeberlen}}]{whh}
\bibinfo{author}{\bibfnamefont{J.}~\bibnamefont{Waugh}},
  \bibinfo{author}{\bibfnamefont{L.}~\bibnamefont{Huber}}, \bibnamefont{and}
  \bibinfo{author}{\bibfnamefont{U.}~\bibnamefont{Haeberlen}},
  \bibinfo{journal}{Phys. Rev. Lett.} \textbf{\bibinfo{volume}{20}},
  \bibinfo{pages}{180} (\bibinfo{year}{1968}).

\bibitem[{\citenamefont{Vandersypen and Chuang}(2004)}]{vandersypen04}
\bibinfo{author}{\bibfnamefont{L.~M.} \bibnamefont{Vandersypen}}
  \bibnamefont{and} \bibinfo{author}{\bibfnamefont{I.~L.}
  \bibnamefont{Chuang}}, \bibinfo{journal}{Rev. Mod. Phys.}
  \textbf{\bibinfo{volume}{76}}, \bibinfo{pages}{1037} (\bibinfo{year}{2004}).

\bibitem[{\citenamefont{Gaudin}(1976)}]{gaudin76}
\bibinfo{author}{\bibfnamefont{M.}~\bibnamefont{Gaudin}}, \bibinfo{journal}{J.
  de Physique} \textbf{\bibinfo{volume}{37}}, \bibinfo{pages}{1087}
  (\bibinfo{year}{1976}).

\bibitem{khaled} Kh. Al--Hassanieh, V. V. Dobrovitski, E. Dagotto, and B. N. Harmon, 
  cond-mat/0511681.

\bibitem{qcqi} M. A. Nielsen and I. L. Chuang, 
  {\it Quantum Computations and Quantum Information\/} (Cambridge University
  Press, Cambridge, 2002).


\bibitem{imamoglu03}
A. Imamoglu~{\em et al.}, Phys. Rev. Lett. \textbf{91}, 017402 (2003).


\bibitem{klauser05}
D. Klauser, W. A. Coish, and D. Loss, e-print: cond-mat/0510177 (2005).


\end{thebibliography}


\end{document}